\documentclass[11pt,nofootinbib,notitlepage,superscriptaddress]{revtex4-1}
\pdfoutput=1

\usepackage{gensymb}
\usepackage{array} 
\usepackage{amssymb}
\usepackage{color}
\usepackage{graphicx}
\usepackage{dcolumn}
\usepackage{bbm}
\usepackage{amsmath}
\usepackage{amsfonts}
\usepackage{wasysym}
\usepackage[pdftex,bookmarks,plainpages=false,pdfpagelabels]{hyperref}

\usepackage{slashed}

\usepackage{url}

\usepackage{booktabs}

\usepackage{float}

\renewcommand{\log}[1]{\mathrm{log}\left(#1\right)}
\newcommand{\tr}[1]{\mathrm{tr}\left(#1\right)}

\newcommand{\dx}{\mathrm{d}x}
\newcommand{\dt}{\mathrm{d}t}
\newcommand{\dr}{\mathrm{d}r}
\newcommand{\dOmega}{\mathrm{d}\Omega}

\begin{document}
\title{Degravitation of the Cosmological Constant in Bigravity}

\author{Moritz Platscher}
\email{moritz.platscher@mpi-hd.mpg.de}
\affiliation{Particle and Astroparticle Physics Division, Max-Planck-Institut f\"ur Kernphysik, \\ Saupfercheckweg 1, 69117 Heidelberg, Germany}
 
\author{Juri Smirnov}
\email{juri.smirnov@mpi-hd.mpg.de}
\affiliation{Particle and Astroparticle Physics Division, Max-Planck-Institut f\"ur Kernphysik, \\ Saupfercheckweg 1, 69117 Heidelberg, Germany}
\affiliation{INFN divisione di Firenze, Dipartimento di Fisica, Università di Firenze,\\
Via Sansone 1, 50019 Sesto Fiorentino, Florence, Italy} 

\begin{abstract}
In this article the phenomenon of degravitation of the cosmological constant is studied in the framework of bigravity. It is demonstrated that despite a sizable value of the cosmological constant its gravitational effect can be only mild. The bigravity framework is chosen for this demonstration as it leads to a consistent, ghost-free theory of massive gravity.  We show that degravitation takes place in the limit where the physical graviton is dominantly  a gauge invariant metric combination.  We  present and discuss several phenomenological consequences expected in this  regime.
\end{abstract}

\thispagestyle{empty}
\maketitle

\section{Introduction}
The nature of the cosmological constant (CC) or vacuum energy poses a great challenge since its introduction by Einstein in General Relativity (GR)~\cite{Einstein:1916vd}.  After experiments confirmed the accelerated expansion of our universe \cite{Perlmutter:1998np}, it was realized that the vacuum energy seems to have a tiny value compared to  all known scales in particle physics \cite{Weinberg:1988cp}. This fact is particularly surprising as it is not protected by any symmetry and is highly sensitive to quantum corrections: any massive particle leads to a loop correction to the value of the vacuum energy, where its mass enters to the fourth power \cite{Martin:2012bt}.  A large number of attempts were made to explain the smallness of the vacuum energy, for example by symmetry arguments~\cite{Wetterich:1987fm}. Furthermore, anthropic arguments for a small value of the CC have been used in the context of the multivese hypothesis \cite{Weinberg:1987dv}. However, a very interesting alternative was suggested in Refs.~\cite{ArkaniHamed:2002fu, Dvali:2002pe, Dvali:2007kt}, where the fundamental idea relies on the possibility that the vacuum energy could actually not have a small value, but gravitate only very mildly. This is possible in theories where four dimensional gravity is mediated by an effectively massive graviton. In Ref.~\cite{Dvali:2007kt} the analogy to the Higgs phase of electromagnetism is used, where due to non-linear interactions of the goldstone fields the field strength induced by a space-time uniform source is damped and the vacuum is de-electrified. In analogy, a toy system modeling gravity is discussed, where it is suggested that the vacuum energy can be degravitated. A following study of the effect in a non-linear extension of Fierz-Pauli massive gravity suggested that cosmic acceleration can be set by the size of the graviton mass scale \cite{deRham:2010tw}.   

In this work we demonstrate the degravitation of the CC in the framework of bigravity. It has recently been proven~\cite{deRham:2010ik,deRham:2010kj,deRham:2011rn,Hassan:2011hr,Hassan:2011vm,Hassan:2011tf,Comelli:2012vz,Deffayet:2012nr,Deffayet:2012zc} that the construction of massive gravity from a bimetric theory suggested in~\cite{Hassan:2011zd,Hassan:2011ea} is free of the Boulware-Deser ghost~\cite{Boulware:1973my}. Thus, we choose this framework as a model of consistent massive gravity to study the degravitation phenomenon. We define a class of models within the bigravity framework in which it can be explicitly shown that above a critical scale $r_V$, the gravitational effect of the CC is suppressed. The suppression of the gravitational effect of the CC in the limit of a dominantly massive gravitational force mediator is the main result of this work and is consistent with the result obatined in linear Fierz-Pauli massive gravity. On the one hand, this is to be expected, since we enter precisely the regime where degravitation is expected to occur. On the other hand, it is known from theories with extra dimensions, which effectively describe massive gravity in 4D, that degravitation may not be fully decoupling the CC~\cite{Dvali:2007kt, deRham:2014zqa}.  Our study focuses on a regime of bigravity, which is close to the massive gravity limit. We are intrigued by the fact that while the near General Relativity limit is well understood \cite{Hassan:2012wr} the opposite regime is not equally well studied. 

This paper is organized in the following way. In Sec.~\ref{sect:Massivegravity}  we briefly describe the bigravity framework, in Sec.~\ref{sect:Solutions} the static spherically symmetric solutions in two different regimes are presented in Subsections \ref{subsect:Linear} and \ref{subsect:Nonlinear}. The phenomenon of degravitation and additional effects are discussed in Sec.~\ref{sect:Discussion}, and our results are summarized Sec.~\ref{sect:Summary}.

\section{A Consistent Massive gravity Framework \label{sect:Massivegravity}}
The bigravity framework is based on the introduction of a second metric  $f$, which does not couple directly to the matter fields, but is connected with the usual metric  $g$ via an interaction term, explicitly constructed in~\cite{deRham:2010kj}. It is a non-linear realization of the St\"uckelberg mechanism which is well-known for the linear theory due to Fierz and Pauli~\cite{Fierz:1939zz,Fierz:1939ix}.  This construction is designed in such a way that no ghost instabilities arise, even at the non-linear level~\cite{deRham:2011rn}. On the linearized level it can be shown that two graviton modes are present. A massless and a massive mode. The physical graviton\footnote{We call the mode which originates from the metric coupled to matter the physical mode. Note that in principle, there could be other forms of coupling to matter. However, as studied in ~\cite{deRham:2014naa} those can lead to the re-introduction of the Boulware--Deser ghost or to a low cut-off scale.} is a superposition of these modes and can be either dominantly massive or massless, depending on the parameter choice of the corresponding mixing angle. In the limit of zero admixture of the massive mode, Einstein GR is consistently recovered, see~\cite{Hinterbichler:2011tt,deRham:2014zqa, Schmidt-May:2015vnx} for some recent reviews.  

The bigravity action for the two tensor fields $g$ and $f$ is given by
\begin{equation}\label{eq:action}
  \begin{aligned}
    S_\mathrm{bi} = &  \int \mathrm{d}^4x \left\lbrace\frac{M_g^2}{2} \sqrt{-|g|} R_g + \frac{M_f^2}{2}  \sqrt{-|f|} R_f +\right. \\ 
   	& \left.+ m^2 M_\text{eff}^2 \sqrt{-|g|} \sum_{n=1}^4 \beta_n e_n(\mathbb{X}) + \sqrt{-|g|} \left( \mathcal{L}_\textrm{matter}  + M_g^2 \Lambda \right)  \right\rbrace.
  \end{aligned}
\end{equation}
Here, $M_g$ is the Planck scale for the $g$ metric, $M_f$ the Planck scale for the $f$ metric, $M_\text{eff}^2 = \left( \frac{1}{M_g^2}+  \frac{1}{M_f^2}\right)^{-1}$,  and $\mathbb{X}$ is defined as $\mathbb{X}^\mu_\alpha \mathbb{X}^\alpha_\nu = g^{\mu\alpha}f_{\alpha\nu}$. Note that, while the vacuum energy $\Lambda$ of the metric $g$ is kept explicitly, the corresponding CC of $f$ is contained in the interaction proportional to the graviton mass. This distinction is made to emphasize that we view $\Lambda$ as the source of the gravitational field, which is renormalized by loops of matter fields coupling only to  $g$.\footnote{The general expression for the vacuum energy as induced by the matter sector reads~\cite{Martin:2012bt,Koksma:2011cq} $\rho_\text{vac} = \sum_i n_i\frac{m_i^4}{64\pi^2}\log{\frac{m_i^2}{\mu_R^2}}+\rho_\text{vac}^\text{EWPT} + \rho_\text{vac}^\text{QCD}$, where $m_i$ are the masses of the matter fields, $\mu_R$ is the renormalization scale, and the $n_i$ count the degrees of freedom and the bosonic/fermionic nature of the field. Note that these contributions are naturally of the order of the SM Higgs field's vacuum expectation value and therefore typically much larger than the graviton mass $m$. Instead of including this contribution in $\beta_0$, in which case one would expect $\beta_0\gg1$, unless fine-tuning is accepted, we include the contribution of the graviton mass to the vacuum energy $\rho_\text{vac}$ and keep $\beta_0 \sim \mathcal{O}(1)$. Degravitation is then necessary to explain the observed, seemingly small value of $\Lambda$.} The fact that matter only couples to the metric $g$ is required by demanding that the theory is ghost free at all scales~\cite{deRham:2014naa, Hassan:2014gta, Schmidt-May:2014xla}, which can be used as a principle for constructing the action.

By varying the action with respect to $g$ and $f$, we obtain two sets of Einstein equations:
\begin{subequations}\label{eq:Einstein}
  \begin{align} 
   &  G(g)_{\mu\nu} + m^2 \sin^2(\theta)\sum_{n=1}^3 \beta_n V^{(n)}(g)_{\mu\nu} = 8 \pi G_N T_{\mu\nu}+ \Lambda g_{\mu\nu},\\
 &    G(f)_{\mu\nu} + m^2  \cos^2(\theta)  \sum_{n=1}^4 \sqrt{\frac{|g|}{|f|}} \beta_n V^{(n)}(f)_{\mu\nu} = 0, 
  \end{align}
\end{subequations}
where $  \sin^2(\theta) = \frac{M_\text{eff}^2}{M_g^2} $, $  \cos^2(\theta) = \frac{M_\text{eff}^2}{M_f^2} $ , and $8 \pi G_N \equiv M_g^{-2}$ is the relation between Newton's constant and the (reduced) Planck mass for $g$. Finally, the interaction or mass terms $V(g/f)$ follow from the variation of the polynomials $e_n$. For more details on the structure of the interaction terms we refer the reader to Appendix ~\ref{App:EinsteinEqs}.

In the following, we will often make use of the assumption that $\beta_4=0$, which usually simplifies the equations significantly. Albeit this assumption, we can directly read off Eqs.~\eqref{eq:Einstein} that this has no effect on the solutions in the limit where the massive mode is dominantly coupling to matter, $\cos \theta \to 0$. In this limit, which corresponds to the limit $M_f \to \infty$, the hidden $f$ metric is fully decoupled from the theory and approaches a static vacuum solution given by $G(f)_{\mu\nu} = 0$, which is \emph{independent} of the value of $\beta_4$. Since $g$ has no interactions proportional to $\beta_4$, we may conclude that in the limit of a dominantly massive graviton coupling to matter, the value of $\beta_4$ can be chosen freely.

In addition, energy-conservation must be ensured by demanding the vanishing of the covariant derivative of Eq.~\eqref{eq:Einstein}:
\begin{equation} \label{eq:Bianchi}
  \nabla(g)_\mu {G(g)^\mu}_\nu = \nabla(g)_\mu {T^\mu}_\nu = 0 \ \Rightarrow \ \nabla(g)_\mu {V^{(n)}(g)^\mu}_\nu = 0.
\end{equation}
Equivalently, we find for the $f$-metric $\nabla(f)_\mu \left(\sqrt{\frac{|g|}{|f|}} {V^{(n)}(f)^\mu}_\nu\right) = 0$. These additional equations are known as Bianchi constraints. However, the equations obtained for $g$ and $f$ are in general \emph{not} independent. 

In the linearized regime of the theory, it can be demonstrated that two spin-2 modes, a massive and a massless one are propagated in the theory.  In the parametrization chosen here, Einstein GR with a  dominantly massless mediator is recovered for $\theta \rightarrow 0$, and the opposite limit, $\theta \rightarrow \pi/2$, leads to  a dominantly massive  physical graviton. A central point is that the massive mode, which is proportional to the combination $\delta g_{\mu \nu} - \delta f_{\mu \nu} $ is invariant under diffeomorphisms acting in the same way on $g$ and  $f$. This makes the massive perturbation a gauge invariant observable~\cite{Giesel:2007wi}.  We are  tempted to  embrace the principle of gauge invariance as the determining principle for the construction of physically observable quantities. Already at this stage of the analysis one can provide a physical argument that, in the limit that this combination is dominantly coupling to matter, any space-time uniform source must decouple, as otherwise its presence would lead to an unbounded growth of a quantity which is a gauge invariant physical observable, thus leading to a paradox~\cite{Dvali:2007kt}.  We will now demonstrate this behavior explicitly in a spherically symmetric system.

\section{Static Spherically symmetric solution \label{sect:Solutions}}
We will study a spherically symmetric system with a massive source at the center and containing a space-time uniform energy density, the CC. Since we are interested in the regime far outside the localized source, we choose an ansatz for the metric, which is convenient for the weak field expansion:
\begin{equation}
  \begin{aligned}
    g_{\mu\nu} \dx^\mu\dx^\nu &= - e^{\nu_1(r)} \dt^2 + e^{\lambda_1(r)} \dr^2 + r^2 \dOmega^2,\\
    f_{\mu\nu} \dx^\mu\dx^\nu &= - e^{\nu_2(r)} \dt^2 + e^{\lambda_2(r)} {(r + r \mu(r))^\prime}^2\dr^2 + (r + r \mu(r))^2 \dOmega^2.\\
  \end{aligned}
\end{equation}
This exponential ansatz is inspired by Ref.~\cite{Babichev:2013pfa} and the exponentials are expanded to leading order in small potential functions $\nu_i$ and $\lambda_i$. The function $\mu$ can be thought of as a relative ``twist'' in the metrics and is a measure of non-linearities induced by the interaction of the metrics.~\footnote{See e.g.~\cite{Babichev:2015xha} for a more general treatment of black hole solutions in bigravity.}

\subsection{The Linear Regime \label{subsect:Linear} }
 
 We assume that at large distances from the source the non-linearities are subdominant and expand the equations assuming  $\mu \ll 1$. 
 
The resulting Einstein equations and Bianchi constraints can be solved exactly, as demonstrated in Appendix \ref{app:Linear}. The respective potential functions read 
\begin{subequations}
  \begin{align}
  \label{Eqn:LinearPotentials}
    \nu_1 (r) =& -\left[\frac{C_1}{r} + \frac{r^2}{3} \cos^2(\theta)\Lambda_\mathrm{eff}\right]  + \sin^2(\theta) \left(\frac{ C_2\, e^{-m r \sqrt{\alpha_1}}}{r}-\frac{2 \widetilde{\Lambda}}{m^2 \alpha_1}  \right) + C_3, \\
    \lambda_1 (r)  =& \frac{C_1}{r} + \frac{r^2}{3} \cos^2(\theta)\Lambda_\mathrm{eff} - \sin^2(\theta) \,\frac{ C_2 e^{-m r \sqrt{\alpha_1}}\left[1 + m r \sqrt{\alpha_1}\right]}{2 r }, \\
    \nu_2 (r) =& -\left[\frac{C_1}{r} + \frac{r^2}{3}\cos^2(\theta) \Lambda_\mathrm{eff}\right] - \cos(\theta)^2  \left(\frac{ C_2\, e^{-m r \sqrt{\alpha_1}}}{r}-\frac{2 \widetilde{\Lambda}}{m^2 \alpha_1}  \right) + C_3, \\
    \lambda_2 (r) =& \frac{C_1}{r} + \frac{r^2}{3} \cos^2(\theta) \Lambda_\mathrm{eff} + \cos(\theta)^2 \frac{ C_2\, e^{-m r \sqrt{\alpha_1}}\left[1+ m r \sqrt{\alpha_1}\right]}{2 r },
  \end{align}
\end{subequations}
with undetermined integration constants $C_{1,2,3,4},\widetilde{\Lambda}$. With the constants  $\alpha_1 \equiv \beta_1 + 2\beta_2 + \beta_3$, $\alpha_2 \equiv   3\beta_1 + 3\beta_2 + \beta_3$, and $\alpha_3 \equiv  \beta_2 + 2 \beta_3$ the potentials have the structure of the Schwarzschild-de~Sitter solution with a superposition of a Yukawa potential and the induced effective vacuum energy $\Lambda_\mathrm{eff} \equiv \left[\Lambda + m^2 \sin^2(\theta)(\alpha_1 + \alpha_2+\alpha_3)\right]$. We note that already at this stage it is obvious that in the limit of the physical graviton being dominantly the massive mode, i.e.\ $\cos(\theta) \rightarrow 0$, the terms in the potential functions proportional to $r^2$ are suppressed. This demonstrates that the strength of the space-time uniform source $\Lambda$ does not have any physical effect in this limit. This was to be expected, as this behavior is known from the linear theory of massive Fierz-Pauli theory. The effects inherent to bigravity will become apparent when non-linearities are taken into account.

The function $\mu$ can be obtained by solving Eq.~\eqref{eq:BianchiFintegrated}:
\begin{equation}
  \mu(r) = \frac{C_2\,e^{-m r \sqrt{\alpha_1}}\left[1+ mr \sqrt{\alpha_1} + m^2r^2\alpha_1 \right]}{4m^2r^3\alpha_1} +\frac{m^2\alpha_1 C_0-2\widetilde{\Lambda}}{6 m^2 \alpha_1} + \frac{C_4}{r^3}\,.
\end{equation}
It is interesting to observe that the form of the metric twist measuring the non-linear effect is independent of the graviton mode mixing. A very important question which needs to be addressed is the range of scales at which this approximation is valid. This can be read of the function $\mu(r)$ provided the integration constants can be determined. Thus to obtain those, we will solve the non-linear equations for radii smaller than the critical radius we call $r_V$ in the next section  and perform a matching of the solutions.

\subsection{The Non-Linear Regime \label{subsect:Nonlinear}}

We are now interested in the form of the solution inside the so-called Vainshtein radius  $r_V$~\cite{Vainshtein:1972sx}. This is the scale below which non-linearities are supposed to become relevant, resolving many of the apparent issues related to massive gravity~\cite{Babichev:2009us,Babichev:2009jt,Babichev:2010jd,Alberte:2010it}. The equations of motion can be simplified under the additional assumption that $r_V \ll \lambda_g \equiv 1/( \sqrt{\alpha_1} m )$, which turns out to be equivalent to the condition that the Schwarzschild radius is smaller than the graviton's Compton wavelength i.e. $r_S \ll \lambda_g$. As we discuss in more detail in Appendix~\ref{app:beta4}, a non-zero value of $\beta_4$ only leads to a redefinition of parameters. Therefore, we can choose it to be  $\beta_4 = 0$ as it greatly simplifies the form of the solutions and helps to display their general features in a compact form. The resulting non-linear equations are
\begin{equation}
\label{Eqn:BianchiNL} 
    \frac{2 \left(r + r\mu(r)\right)^\prime \left(\lambda_1(r)-\lambda_2(r)\right)}{ \left(r + r\mu(r)\right)^\prime
   \nu_1^\prime(r)-\nu_2^\prime(r)}
    = 
   r \frac{\alpha_1+2 \alpha_4\, \mu (r)+(\alpha_3 - \alpha_4) \mu (r)^2}{\alpha_1+\alpha_4\, \mu (r)}\,,
\end{equation} 
where $\alpha_4  \equiv  \beta_2 + \beta_3$.

Choosing $\alpha_3 = 0$ and $\alpha_4=-\alpha_1$, also for the sake of brevity, one obtains very compact solutions inside the Vainshtein radius $r_V$ as further discussed in Appendix \ref{app:Nonlinear}. The solutions outside a source are:\footnote{Inside a spherical matter distribution of radius $R_0$ the $g$ metric potential is $\nu(r)= - \frac{\rho r^2}{3 M_g^2} - \frac{r^2}{3} \Lambda_{\rm eff}$, while the $f$ metric potential stays unaffected. Matching of this inner and outer solutions fixes the integration constant $r_S = 2 M G_N$, where $M$ is the total mass of the source.}:
\begin{subequations}
\begin{align}
\label{Eqn:NLPotentials}
\nu_1 (r) &   = -  \lambda_1(r)  = - \frac{r_S}{r} - \frac{r^2}{3} \left[  \Lambda +  m^2 \sin^2(\theta ) \left(  \alpha_1 +  \alpha _2 \right) \right]=- \frac{r_S}{r} - \frac{r^2}{3} \Lambda_\text{eff} \,, \\
  \nu_2(r) &  = - \lambda_2(r)  =  -\frac{2}{3} \, \alpha _1  m^2 r^2 \cos^2(\theta )\,.
\end{align} 
\end{subequations}
We observe that the potential of the $g$ metric which couples to matter reproduces the GR Schwarzschild-de Sitter solution with an induced CC with contributions from the bare CC and the graviton mass. In the limit of $\theta \rightarrow 0$ this expression reproduces standard GR.\footnote{Note that this limit is continuous in the sense that all massive d.o.f.\ decouple~\cite{Baccetti:2012bk}. This is different from the limit $m \to 0$ where the so-called van Dam-Veltman-Zakharov discontinuity~\cite{vanDam:1970vg,Zakharov:1970cc} occurs: for a given source, the helicity-0 mode does not decouple when $m\to 0$, as can be seen from Eq.~\eqref{Eq:FullPotential}.}

The linear and non-linear solutions can be matched at $r_V$, as shown in Appendix \ref{app:Matching}, leading to 
\begin{equation}
\label{Eqn:MatchedConsts}
\begin{gathered}
C_0 \Rightarrow \widetilde{\Lambda} = 0, \quad C_1 = r_S\, \cos^2(\theta) \left( 1 + \frac{2}{3} \sin^2(\theta)\right),\quad  C_2 = -2\, r_S \frac{\Lambda +m^2 \alpha_1 + m^2 \sin^2(\theta) (3\alpha_1 + \alpha_2)}{3 m^2\alpha_1}  ,\\
C_3 = - \cos^2(\theta) m \sqrt{\alpha_1} C_2,\quad C_4 = r_S \frac{5 \Lambda + 3 \alpha_1 m^2 + m^2 \sin ^2(\theta ) (7 \alpha_1+ 5 \alpha_2)}{6\alpha_1^2 m^4},
\end{gathered}
\end{equation}
with $r_S$ being the Schwarzschild radius of the central matter distribution. 

As mentioned before, the  linearity assumption $\mu \ll 1 $ is violated as $r \to 0$. With the matched coefficients it is clear under which condition the function $\mu$ is small: we can read of that if 
\begin{align}
r \gg \left(\frac{r_S}{m^2 \,\alpha_1}\right)^{\frac{1}{3}} \equiv r_V\,,
\end{align}
the assumption $\mu \ll 1  $ is justified and we are indeed in the linear regime.  This condition defines the scale $r_V$ and justifies our initial expansion in $\mu(r) \ll 1$ \emph{a posteriori}. 

After the matching we obtain the following physical gravitational potential:
\begin{equation}
\label{Eq:FullPotential}
\nu_1(r) =
\begin{cases}
 - \frac{r_S}{r} -r^2 \, \frac{\Lambda_\mathrm{eff} }{3}  & r \ll r_V
 \\
- \frac{r_S }{r} \left[ \cos^2(\theta)\left(1+\frac{2}{3}\sin^2(\theta)\right) + \frac{2}{3} \sin^2(\theta) e^{- m \, \sqrt{\alpha _1}\, r}\left(1 + 2 \sin^2(\theta) + \frac{\Lambda_\text{eff}}{m^2\alpha_1} \right)\right]- & \\
\hfill -r^2\cos^2(\theta) \, \frac{\Lambda_\mathrm{eff} }{3} & r \gg r_V
\,. 
\end{cases}
\end{equation}
The derivation of this potential is the main result of this work. It demonstrates that inside the Vainshtein radius we have the known Schwarzschild-de Sitter solution with the induced CC $\Lambda_\mathrm{eff} = \Lambda + m^2 \sin^2(\theta) \left(\alpha_1 + \alpha_2\right)$. Since inside the Vainshtein radius, the longitudinal modes become strongly coupled~\cite{ArkaniHamed:2002sp}, no screening effect is seen.  Outside the Vainshtein radius, the induced CC is $\Lambda_\mathrm{eff}^\text{outer} = \cos(\theta)^2 \Lambda_\mathrm{eff}$ which is decoupled in the massive limit $\cos(\theta) \rightarrow 0$. Also, in the massive limit the outer potential is of Yukawa type. The induced CC has a contribution from the bare CC and the one induced by the graviton mass effect. This suppression in the massive gravity limit was expected from the study of linear Fierz-Pauli massive gravity. However, our result shows explicitly how the non-linear effects in a full theory shut off the screening below the critical scale which is set by the Vainshtein radius. 

At this point, we may finally describe how degravitation works in the bigravity framework. Taking the limit of a purely massive graviton coupling to matter, i.e. $\sin\theta = 1$ and $\cos \theta = 0$, we see that the potential inside the Vainshtein radius is purely newtonian \emph{with} an effective CC, while outside $r_V$ we have a Yukawa potential $e^{-m_g r}/r$ with the effective CC \emph{decoupled}. Therefore, we see that we have found the scale above which the vacuum energy density is degravitated. This is a remarkable result and to our knowledge has not been pointed out in the literature so far within this framework and represents a genuine realization of the mechanism described in Ref.~\cite{Dvali:2007kt} within a UV complete theory of massive gravity. An important fact becomes obvious at this point. We see that the exact limit $\cos(\theta) = 0$, the massive gravity regime, can not lead to a consistent cosmology, as no late-time acceleration would be possible. Thus, the physically relevant theory can only have a small, but finite mixing $\cos(\theta) \ll 1 $. We comment on this fact in more detail in the next section.

\section{Degravitation and Induced Effects \label{sect:Discussion}}

We will now present some phenomenological implications of our findings. As discussed in the introduction, the theoretical expectation for the bare value of $\Lambda$ is large due to effects of vacuum loops and phase transitions in the universe. 

From the form of Eq.~\eqref{Eq:FullPotential}, we observe that there are two effects at work in massive gravity. On the one hand we demonstrate how degravitation as suggested by Dvali et al.~\cite{Dvali:2007kt} works in the linear regime. This phenomenon is manifest in the limit $\cos(\theta) \rightarrow 0$, which corresponds to the dominance of the massive mode. In this limit the potential outside the Vainshtein radius is a pure Yukawa potential and the space-time uniform contribution of the CC decouples.  Thus in this long range effect there is no fine tuning involved as an arbitrary value of $\Lambda_\mathrm{eff}$ gravitates arbitrarily mildly, as long as $\cos(\theta)$ is small enough. Setting the $\cos(\theta)$ to a small value could be seen as a fine tuning as well, however there are good reasons to assume that physical arguments as gauge invariance of the physical graviton --~the key principle to working degravitation as we have argued~-- favor this choice. 

On the other hand, inside the Vainshtein radius, there is always a remaining effect of the CC. And in order to keep $\Lambda_\mathrm{eff}$ in agreement with solar system observations fine tuning is unavoidable and we will comment on that shortly.

At this point it is also important to comment on a curious fact. In practice, whenever limits on the graviton mass or the value of the CC are set from solar system observations, each effect is treated separately. However, we showed that those two effects are intrinsically intertwined in such a way that they cannot be  kept apart inside the Vainshtein radius. Therefore, on the one hand the limits we obtain from planetary orbits are limits on the combination $\Lambda_\text{eff}$. On the other hand, the distance scale we consider those observations at sets a limit on the Vainshtein radius and hence on the graviton mass. 

The  vast majority of known objects in the solar system are inside the Kuiper belt, which has a 50 AU radius. We observe that in the degravitating regime, it is unacceptable that the Vainshtein radius of the solar system is smaller than its size. As the gravitational law would change significantly on very well studied scales, contradicting observations [see  Eqn. \eqref{Eq:FullPotential}]. If the Vainshtein radius of the solar system is taken to be larger than the radius of the Kuiper belt, we obtain a graviton mass bound of $m_g = m\, \sqrt{\alpha_1} < 5 \cdot 10^{-25} \text{ eV}$. This bound is a conservative estimate and is slightly stronger than the solar system bound reported in \cite{deRham:2016nuf}, where a generic expanded Yukawa potential has been considered.  However, both bounds are significantly lower than the upper mass bound obtained from cluster observations, $m_g < 10^{-29} \text{ eV}$ \cite{deRham:2016nuf}.

Having established that in the degravitating regime the Vainshtein radius has to be outside the solar system, we can set a limit on $\Lambda_\text{eff}$. The limit which can be obtained on the value of $\Lambda_\text{eff} = (\rho_\text{induced}^\text{vac})/M_p^2$ from the perihelion rotation of Mercury is $\rho_\text{induced}^\text{vac}< \left( 14 \text{ eV} \right)^4$ \cite{Martin:2012bt}. This bound is sixteen orders of magnitude larger than the value obtained from late time acceleration $\rho_\text{asymptotic}^\text{vac} \approx \left( 1.8 \cdot 10^{-3} \text{ eV} \right)^4$. Nevertheless, it is still significantly smaller than, for example the value one expects to be added to the CC from the QCD phase-transition. Thus, there remains some fine tuning, which is however sixteen orders of magnitude milder than in GR since we can saturate the bound locally, but still obtain an asymptotic vacuum energy density in agreement with late time acceleration.  

At the same time, assuming that we are in the degravitation regime, a lower bound on the mixing angle can be obtained. Assuming that $\Lambda_\text{eff}$ saturates the solar system bounds inside the Vainshtein radius and knowing that the value of the asymptotic CC on the largest scales leads to  $\rho_\text{asymptotic}^\text{vac} \approx \left(1.8\cdot 10^{-3}\text{ eV}\right)^4$, we can infer that $\cos(\theta)> 1.8 \cdot 10^{-8}$. This is an interesting result, which shows that exact massive gravity does not describe our universe. A fact which deserves aditional attention is that in bigravity, in a cosmological solution, the physical Planck mass can be different from its local value \cite{Akrami:2015qga, vonStrauss:2011mq}. This investigation however is postponed to future work.

If indeed  our universe is close to the degravitated regime, we see from Eq.~\eqref{Eq:FullPotential} that unavoidably the gravitational potential is modified at the transition around the Vainshtein radius and is enhanced by a leading order factor 
\begin{align}
\label{Eqn:PotentialDeviation}
C \approx \cos^2(\theta) \left( 1+ \frac{2}{3}¸\sin^2(\theta)\right)+ \frac{2 \sin^2(\theta)}{3} \left(1 + 2 \sin ^2(\theta ) + \frac{\Lambda _{\text{eff}}}{\alpha _1 m^2} \right) \approx \frac{2}{3}\left( 3 + \frac{\Lambda _{\text{eff}}}{\alpha _1 m^2} \right)\,.
\end{align}
This is due to the fact that inside the Vainshtein radius the longitudinal modes of the gravitons are strongly coupled and conspire to reproduce the GR predictions. However, at the transition to the linear regime those modes become weakly coupled and lead to an enhanced force. The enhancement could lead to observable deviations from Einstein gravity at scales larger than the Vainshtein radius of a given system. For example for the Andromeda galaxy with $1.5\cdot 10^{12}$ solar masses, and given the cluster bound on the graviton mass, the critical radius  $r_V$ would be larger than $4$ kiloparsecs. Thus, deviations from pure GR behavior are expected to be only observable in large systems as galaxies or clusters where the effects appear above the kiloparsec length scale. 

To test the degravitation hypothesis gravitational lensing measurements might be more sensitive. The deflection angle outside the Vainshtein radius is approximately given by 
\begin{align}
\Delta \phi \approx \frac{r_S \left(3 \left(\sin ^2(\theta ) \left(\Lambda _{\text{eff}}+\alpha _1 m^2\right)+2 \alpha _1 m^2\right)-2 \alpha _1
   m^2 \sin ^2(\theta ) \cos ^2(\theta )\right)}{3 \alpha _1 m^2 r_i}\,,
\end{align}
where  $r_i$ is the distance at which the light is passing by the source.  The expression  has the limiting behaviors 
\begin{subequations}\label{eq:deflectionAngle}
\begin{align}
& \Delta \phi \rightarrow \frac{2 r_S}{r_i} &&\text{  for  } \theta \rightarrow 0 \,\text{, GR limit} \\
& \Delta \phi \rightarrow \frac{r_S}{r_i} \left(3 + \frac{\Lambda _{\text{eff}}}{\alpha_1 m^2} \right) &&\text{  for  } \theta \rightarrow \frac{\pi}{2}\,\text{, massive gravity limit} \,.
\end{align}
\end{subequations}
In conclusion, we find that in the regime leading to degravitation substantial deviations from the GR light deflection are expected at large distance scales. We leave a detailed phenomenological survey for future work.

\section{Summary \label{sect:Summary}}

In this article we have demonstrated the effect of degravitation in a concrete model of massive gravity. We have shown that, in the limit where the physical gravity mediator is dominantly a massive, gauge invariant mode, the effect of a space-time uniform source coupled to the physical metric is suppressed.  We emphasize that in the degravitating regime, limits on the graviton mass have to be set in a different way, as the gravitational law changes substantially at the Vainshtein radius, a critical length scale we defined in the theory.  Furthermore, we comment on the possibility to test the degravitation hypothesis by measuring  deviations of the gravitational potential and light deflection at scales larger than this critical scale. 

\section*{Acknowledgments}
The authors thank Matthias Bartelmann and Angnis Schmidt-May for invaluable discussions and comments on the manuscript. MP is supported by IMPRS-PTFS. 

\appendix

\section{Einstein Equations in Bigravity}\label{App:EinsteinEqs}

Let us briefly discuss the details of obtaining the field equations for the tensor fields $g$ and $f$ from the action~\eqref{eq:action}. There, the $e_n$ are the  elementary symmetric polynomials of the eigenvalues of the matrix $\mathbb{X}$. These can be expressed as
\begin{subequations}\allowdisplaybreaks
  \begin{align}
    e_1 &= \tr{\mathbb{X}},\\
    e_2 &= \frac{1}{2} \left[ \tr{\mathbb{X}}^2 - \tr{\mathbb{X}^2}\right],\\
    e_3 &= \frac{1}{6} \left[ \tr{\mathbb{X}}^3 - 3\, \tr{\mathbb{X}}\tr{\mathbb{X}^2}+2\, \tr{\mathbb{X}^3}\right],\\
    e_4 &= \mathrm{det}(\mathbb{X}).
  \end{align}
\end{subequations}
The variation of the action~\eqref{eq:action} w.r.t.\ $g$ yields the interaction terms
\begin{subequations}
  \begin{align}
    {V^{(1)}(g)^\mu}_\nu =& \tr{\mathbb{X}}{\delta^\mu}_\nu - {\mathbb{X}^\mu}_\nu ,\\
    {V^{(2)}(g)^\mu}_\nu =& {\left(\mathbb{X}^2\right)^\mu}_\nu - \tr{\mathbb{X}} {\mathbb{X}^\mu}_\nu + \frac{{\delta^\mu}_\nu}{2}\left[\tr{\mathbb{X}}^2 - \tr{\mathbb{X}^2}\right],\\
    {V^{(3)}(g)^\mu}_\nu =& -{\left(\mathbb{X}^3\right)^\mu}_\nu + \tr{\mathbb{X}} {\left(\mathbb{X}^2\right)^\mu}_\nu
			  - \frac{1}{2} \left[\tr{\mathbb{X}}^2 - \tr{\mathbb{X}^2}\right]{\mathbb{X}^\mu}_\nu +\nonumber \\
			  & + \frac{{\delta^\mu}_\nu}{6} \left[ \tr{\mathbb{X}}^3 - 3\, \tr{\mathbb{X}}\tr{\mathbb{X}^2}+2\, \tr{\mathbb{X}^3}\right].
  \end{align}
\end{subequations}
The expressions for $V^{(1,2,3)}(f)$ are obtained from these by dropping the parts proportional to ${\delta^\mu}_\nu$ and multiplying by $(-1)$, while for $n=4$, one obtains $V^{(4)}(f) = {\delta^\mu}_\nu$.

\section{Spherically symmetric and static black hole solution}

In this appendix we discuss how to obtain the solution to the classical, weak-field potential given in Eq.~\eqref{Eq:FullPotential}. Our starting point is the ansatz
\begin{equation}
  \begin{aligned}
    g_{\mu\nu} \dx^\mu\dx^\nu &= - e^{\nu_1(r)} \dt^2 + e^{\lambda_1(r)} \dr^2 + r^2 \dOmega^2,\\
    f_{\mu\nu} \dx^\mu\dx^\nu &= - e^{\nu_2(r)} \dt^2 + e^{\lambda_2(r)} {(r + r \mu(r))^\prime}^2\dr^2 + (r + r \mu(r))^2 \dOmega^2,\\
  \end{aligned}
\end{equation}
which we plug into the Einstein Equations derived in the previous appendix.

\subsection{The Linear Regime \label{app:Linear}}

At large distances from the source, one assumes a space-time which is nearly flat, such that in addition, we take $\mu \ll 1$. With this assumption we obtain the following linearized Einstein equations:
\begin{subequations}\allowdisplaybreaks\label{eq:gEinsteinEq}
  \begin{align}
    \frac{\lambda_1}{r^2} + \frac{\lambda_1^\prime}{r} &= \Lambda_g + m^2_g \left[ \frac{1}{2}(\lambda_2-\lambda_1) + \left(3 \mu + r \mu^\prime \right) \right],\label{eq:gEinsteinEq1}\\
    \frac{\lambda_1}{r^2} - \frac{\nu_1^\prime}{r} &= \Lambda_g + m^2_g \left[ \frac{1}{2}(\nu_2-\nu_1) + 2  \mu \right],\label{eq:gEinsteinEq2}\\
    \frac{1}{2} \left( \frac{\lambda_1^\prime}{r} - \frac{\nu_1^\prime}{r} - \nu_1^{\prime\prime}\right) &= \Lambda_g + m^2_g \left[ \frac{1}{2}(\lambda_2-\lambda_1+\nu_2-\nu_1) + \left(2 \mu + r \mu^\prime \right) \right]
  \end{align}
\end{subequations}
with $m^2_g = m^2 \alpha_1 \sin(\theta)^2$ and $\Lambda_g = \Lambda+m^2 \alpha_2 \sin(\theta)^2$ for the $g$ metric.  And similarly for the $f$ metric:
\begin{subequations}\label{eq:fEinsteinEq}
  \begin{align}
    \frac{\lambda_2}{r^2} + \frac{\lambda_2^\prime}{r} &=  \Lambda_f + m^2_f \left[ \frac{1}{2}(\lambda_1-\lambda_2) - \left( 3\mu + r \mu^\prime\right)\right],\label{eq:fEinsteinEq1}\\
    \frac{\lambda_2}{r^2} - \frac{\nu_2^\prime}{r} &=  \Lambda_f + m^2_f \left[ \frac{1}{2}(\nu_1 -\nu_2) - 2\, \mu \right],\label{eq:fEinsteinEq2}\\
    \frac{1}{2} \left( \frac{\lambda_2^\prime}{r} - \frac{\nu_2^\prime}{r} - \nu_2^{\prime\prime}\right) &=  \Lambda_f + m^2_f \left[ \frac{1}{2}(\lambda_1 + \nu_1 - \lambda_2 - \nu_2) -  \left( 2\mu + r \mu^\prime\right)\right].
  \end{align}
\end{subequations}
with $\Lambda_f= m^2 (\alpha_1 + \alpha_3) \cos^2(\theta)$ and $ m^2_f = m^2 \alpha_1 \cos^2(\theta)$. We have defined $\alpha_1 \equiv \beta_1 + 2\beta_2 + \beta_3$, $\alpha_2 \equiv + 3\beta_1 + 3\beta_2 + \beta_3$, and $\alpha_3 \equiv  \beta_2 + 2 \beta_3$ and for the sake of compactness of the expressions fixed $\beta_4=0$. On the one hand, this does not change the physical potentials and implies that the $f$ metric (which does not couple to matter) has only an effective vacuum energy contribution proportional to the graviton mass.

The independent Bianchi constraints read
\begin{subequations}\label{eq:Bianchi}
  \begin{align}
    \lambda^{(-)} - \frac{r}{2}  {\nu^{(-)}}^\prime &=0, \label{eq:BianchiG}\\
 {\lambda^{(-)}}^\prime + {\nu^{(-)}}^\prime - 8\mu^\prime - 2r \mu^{\prime\prime}  &=0\,.\label{eq:BianchiF}
  \end{align}
\end{subequations}
Here, we have introduced the notation $\lambda^{(-)} \equiv \lambda_1 - \lambda_2$ and $\nu^{(-)} \equiv \nu_1 - \nu_2$.  We can integrate~\eqref{eq:BianchiF}, using~\eqref{eq:BianchiG}, and obtain
\begin{equation} \label{eq:BianchiFintegrated}
  \left(r^3\mu\right)^\prime = \frac{r}{4} \left(r^2 \nu^{(-)}\right)^\prime + \frac{C_0\,r^2}{2}
\end{equation}
This may be used to simplify the square brackets in Eqs.~\eqref{eq:gEinsteinEq1} and~\eqref{eq:fEinsteinEq1}:
\begin{equation}
  \mp \frac{1}{2}\lambda^{(-)} \pm \left( 3\mu + r \mu^\prime\right) = \mp \frac{r}{4}{\nu^{(-)}}^\prime \pm \frac{1}{4r}\left( 2r {\nu^{(-)}} + r^2 {\nu^{(-)}}^\prime \right) \pm \frac{C_0}{2}= \pm \frac{1}{2}\nu^{(-)} \pm \frac{C_0}{2}.
\end{equation}
Subtracting Eqs.~\eqref{eq:gEinsteinEq1} and~\eqref{eq:fEinsteinEq1} yields
\begin{equation}
  \frac{1}{r^2} \left( r \lambda^{(-)}\right)^\prime = \frac{1}{r^2} \left( \frac{r^2}{2} {\nu^{(-)}}^\prime\right)^\prime = \widetilde{\Lambda}+\frac{m^2 \alpha_1}{2} \nu^{(-)},
\end{equation}
with $\widetilde{\Lambda} \equiv \Lambda + m^2 \left(\alpha_1 C_0/2 +\sin^2(\theta) \alpha_2 - \cos^2(\theta) \left(\alpha_1 + \alpha_3 \right) \right)$. This equation has the general solution
\begin{subequations}
\begin{align}
  \nu^{(-)}(r) &= - \frac{2 \widetilde{\Lambda}}{m^2 \alpha_1} + \frac{C_2}{r} e^{-m r \sqrt{\alpha_1}},\\
 \Rightarrow \ \lambda^{(-)}(r) &= - \frac{C_2\left[1+m r \sqrt{\alpha_1}\right]}{2r} e^{-m r \sqrt{\alpha_1}},
\end{align}
\end{subequations}
where Eq.~\eqref{eq:BianchiF} has to be used.
Note that here the exponentially divergent branch has not been considered.

Similarly, we can consider the linear combination $\lambda^{(+)} \equiv \cos(\theta)^2 \lambda_1 + \sin^2(\theta) \lambda_2$ which, using Eqs.~\eqref{eq:gEinsteinEq1} and~\eqref{eq:fEinsteinEq1}, obeys the differential equation
\begin{equation}
  \frac{1}{r^2}\left(r\lambda^{(+)}\right)^\prime = \cos(\theta)^2 \left[ \Lambda + m^2 \sin^2(\theta) (\alpha_1 + \alpha_2 + \alpha_3) \right]
\end{equation}
and may be integrated to yield
\begin{equation}\label{eq:lambdaPlusSol}
  \lambda^{(+)}(r) = \frac{C_1}{r} + \frac{r^2}{3} \cos(\theta)^2 \Lambda_\mathrm{eff},
\end{equation}
with $\Lambda_\mathrm{eff} \equiv \left[\Lambda + m^2 \sin^2(\theta)(\alpha_1 + \alpha_2+\alpha_3)\right]$.

Finally, from Eqs.~\eqref{eq:gEinsteinEq2} and~\eqref{eq:fEinsteinEq2}, one may find [$\nu^{(+)} \equiv \cos(\theta)^2\nu_1 + \sin^2(\theta) \nu_2$]
\begin{equation}\label{eq:nuPlusSol}
  \nu^{(+)}(r) = - \left[ \frac{C_1}{r} + \frac{r^2}{3} \cos(\theta)^2 \Lambda_\mathrm{eff} - C_3\right] .
\end{equation}
Thus, we find 
\begin{subequations}
  \label{Eqn:LinearPotentials}
  \begin{align}
    \nu_1 (r) = \nu^{(+)}(r) + \sin^2(\theta)\, \nu^{(-)}(r) =& -\left[\frac{C_1}{r} + \frac{r^2}{3} \cos^2(\theta)\Lambda_\mathrm{eff}\right]  + \sin^2(\theta) \left(\frac{ C_2\, e^{-m r \sqrt{\alpha_1}}}{r}-\frac{2 \widetilde{\Lambda}}{m^2 \alpha_1}  \right) + C_3, \\
    \lambda_1 (r) = \lambda^{(+)}(r) + \sin^2(\theta) \,\lambda^{(-)}(r) =& \frac{C_1}{r} + \frac{r^2}{3} \cos^2(\theta)\Lambda_\mathrm{eff} - \sin^2(\theta) \,\frac{ C_2 e^{-m r \sqrt{\alpha_1}}\left[1 + m r \sqrt{\alpha_1}\right]}{2 r }, \\
    \nu_2 (r) = \nu^{(+)}(r) - \cos(\theta)^2 \nu^{(-)}(r) =& -\left[\frac{C_1}{r} + \frac{r^2}{3}\cos^2(\theta) \Lambda_\mathrm{eff}\right] - \cos(\theta)^2  \left(\frac{ C_2\, e^{-m r \sqrt{\alpha_1}}}{r}-\frac{2 \widetilde{\Lambda}}{m^2 \alpha_1}  \right) + C_3, \\
    \lambda_2 (r) = \lambda^{(+)}(r) - \cos(\theta)^2  \lambda^{(-)}(r) =& \frac{C_1}{r} + \frac{r^2}{3} \cos^2(\theta) \Lambda_\mathrm{eff} + \cos(\theta)^2 \frac{ C_2\, e^{-m r \sqrt{\alpha_1}}\left[1+ m r \sqrt{\alpha_1}\right]}{2 r }.
  \end{align}
\end{subequations}
The function $\mu$ can be obtained by integrating Eq.~\eqref{eq:BianchiFintegrated}:
\begin{equation}
  \mu(r) = \frac{C_2\,e^{-m r \sqrt{\alpha_1}}\left[1+ mr \sqrt{\alpha_1} + m^2r^2\alpha_1 \right]}{4m^2r^3\alpha_1} +\frac{m^2\alpha_1C_0 - 2\widetilde{\Lambda}}{6 m^2 \alpha_1} + \frac{C_4}{r^3}\,. 
\end{equation}

\subsection{The Non-Linear Regime  \label{app:Nonlinear}}

If we assume that $r_V \ll \lambda_g \equiv 1/( \sqrt{\alpha_1} m )$, which is equivalent to the condition that $r_S \ll \lambda_g$, the equations of motion can be integrated assuming that $\beta_4 = 0 $, as it was done in Ref.~\cite{Babichev:2013pfa}. We generalize this procedure to a non vanishing CC and obtain from the Bianchi constraint
\begin{equation}
\label{Eqn:BianchiNLApp} 
    \frac{2 \left(r + r\mu(r)\right)^\prime \left(\lambda_1(r)-\lambda_2(r)\right)}{ \left(r + r\mu(r)\right)^\prime
   \nu_1^\prime(r)-\nu_2^\prime(r)}
    = 
   r \frac{\alpha_1+2 \alpha_4\, \mu (r)+(\alpha_3 - \alpha_4) \mu (r)^2}{\alpha_1+\alpha_4\, \mu (r)},
\end{equation} 
together with the integrated expressions for the potential functions, 
\begin{subequations}\allowdisplaybreaks
\begin{align}
	\nu_1^\prime (r) &   = \frac{r_S}{r^2}-\frac{2}{3}\Lambda_g  r + \frac{1}{3} m^2 r \sin^2(\theta) \mu(r)\left[ -3 \alpha_1 + (\alpha_3-\alpha_4)\mu(r)^2 \right] \,, \\
 	\lambda_1(r)  &  =  \frac{r_S}{r}+\frac{1}{3}\Lambda_g  r^2 + \frac{1}{3} m^2 r^2 \sin^2(\theta) \mu(r)\left[  3 \alpha_1 + 3 \alpha_4 \mu(r) +  (\alpha_3-\alpha_4)\mu(r)^2 \right] \,, \\
 	\nu_2^\prime (r) &  =  -\left[ \frac{2}{3}\Lambda_f\,  r + m^2 r \cos^2(\theta) \mu(r)\left(  \alpha_1 + 2 \alpha_3 + 2 \alpha_3 \mu(r) + (\alpha_3-\alpha_4)\mu(r)^2 \right) \right] \frac{(r+r\mu(r))^\prime}{(1+\mu(r))^2} \,, \\ 
	\lambda_2(r) &  =  \left[\frac{1}{3} \Lambda_f\, r^2+  m^2 \cos^2(\theta)r^2 \mu(r)\left(\alpha_3 + (\alpha_3-\alpha_4)\mu(r)\right) \right]  (1+\mu(r))^{-1}
\end{align} 
\end{subequations}
a seventh order algebraic equation for $\mu(r)$, where $\alpha_4  \equiv  \beta_2 + \beta_3$. Note that the integration constants have been chosen from the continuity condition of the potential functions at the source, in the same way as in \citep{Babichev:2013pfa}.  

The fact that this equation will always lead to constant solutions for $\mu$ is a direct consequence of the Vainshtein mechanism since it would induce non-GR dependencies of the potential functions on $r$. However, inside the Vainshtein radius, only standard GR should be present. The most general solution will be a root of some function, $B(\mu, m, r_S, \alpha_{1,2,3,4}, \theta)=0$. Interestingly, there is always a solution $\mu = 1$, which upon comparing coefficients in $r$, demands that $\alpha_3 = 0$ and $\alpha_4=-\alpha_1$. We will use this solution for illustration here, and refer the reader to Ref.~\cite{Babichev:2013pfa} for a more detailed analysis of these solutions. However, we note that this will merely lead to a redefinition of the physical observables $r_S$ and $\Lambda$ in terms of $\mu = \text{const} \neq 1$. 
 This solution leads to very compact expressions for the potentials inside the Vainshtein radius
\begin{subequations}
\label{Eqn:NLPotentials}
\begin{align}
\nu_1 (r) &   = -  \lambda_1(r)  = - \frac{r_S}{r} - \frac{r^2}{3} \left[  \Lambda +  m^2 \sin^2(\theta ) \left(  \alpha_1 +  \alpha _2 \right) \right]\,, \\
  \nu_2(r) &  = - \lambda_2(r)  =  -\frac{2}{3} \, \alpha _1  m^2 r^2 \cos^2(\theta )\,.
\end{align} 
\end{subequations}

\subsection{Matching of both Regimes  \label{app:Matching}}

The five integration constants $C_{0,1,2,3,4}$ of the general, linearized solution can be fixed by matching to the solution in the non-linear regime at $r= r_V$. The resulting values are:
\begin{equation}
\begin{gathered}
C_0 \Rightarrow \widetilde{\Lambda} = 0, \quad C_1 = r_S\, \cos^2(\theta) \left( 1 + \frac{2}{3} \sin^2(\theta)\right),\quad  C_2 = -2\, r_S \frac{\Lambda +m^2 \alpha_1 + m^2 \sin^2(\theta) (3\alpha_1 + \alpha_2)}{3 m^2\alpha_1}  ,\\
C_3 = - \cos^2(\theta) m \sqrt{\alpha_1} C_2,\quad C_4 = r_S \frac{5 \Lambda + 3 \alpha_1 m^2 + m^2 \sin ^2(\theta ) (7 \alpha_1+ 5 \alpha_2)}{6\alpha_1^2 m^4},
\end{gathered}
\end{equation}
with $r_S$ being the Schwarzschild radius of the central matter distribution. To obtain these expressions, we have assumed that $r_S \ll r_V$ and kept only terms linear in $r_S$. However, the result for $C_1$ is valid independently of this approximation. Our result is the generalization of the special case given in Ref.~\cite{Babichev:2013pfa}. 

\subsection{The effect of $\beta_4\neq0$}\label{app:beta4}
So far, we have argued on intuitive grounds that choosing $\beta_4=0$ has no significant effect on our analysis. We have also verified this statement numerically and find agreement, i.e.\ degravitation occurs irrespective of the value of $\beta_4$. The main effect induced by a $\beta_4 \neq 0$ is in fact a different relation between matter and mass bases. This can be seen as follows: while the $g$-metric Einstein equations~\eqref{eq:gEinsteinEq} are unchanged, the $f$ equations~\eqref{eq:fEinsteinEq} receive an additional contribution on the r.h.s.\ given by
\begin{equation}
	\text{Eqs.~\eqref{eq:fEinsteinEq}} \rightarrow \text{Eqs.~\eqref{eq:fEinsteinEq}} + m^2\cos^2\theta\, \beta_4 \left[ \frac{1}{2}(\lambda_1-\lambda_2) + \frac{1}{2} (\nu_1 - \nu_2) - \left( 3\mu + r \mu^\prime\right) \right].
\end{equation}
Therefore, we can take differences of the $g$ and $f$ equations, as before, and find
\begin{equation}
	\nu^{(-)}(r) = - \frac{2 \hat{\Lambda}}{m^2 \alpha_1} + \frac{C_2}{r} e^{-m r \sqrt{\alpha_1}} \quad
 	\Rightarrow \quad \lambda^{(-)}(r) = - \frac{C_2\left[1+m r \sqrt{\alpha_1}\right]}{2r} e^{-m r \sqrt{\alpha_1}},
\end{equation}
which is formally identical to the previous solutions, but one has to redefine $\widetilde{\Lambda} \to \hat{\Lambda}$, which is a function of $\beta_4$. Qualitatively, the only difference occurs for the ``$(+)$'' functions. While we previously had relations of the sort $\nu^{(+)} \equiv \cos(\theta)^2\nu_1 + \sin^2(\theta) \nu_2$, these are now modified as
\begin{equation}
	\widehat{\nu}^{(+)} \equiv \frac{\alpha_1 + \beta_4}{\alpha_1}\cos(\theta)^2\nu_1 + \sin^2(\theta) \nu_2,
\end{equation}
and similarly for $\lambda^{(+)}$. Clearly, in the case where $\cos \theta \to 0$, $\beta_4$ drops out of this relation and all other equations, confirming our intuitive argument given in Sec.~\ref{sect:Massivegravity}. In the general case, however, this will no longer disentangle the massive and massless modes and the solutions show both types of behavior. However, no new $r$-dependencies are introduced and most importantly, the solutions smoothly approach Eqs.~\eqref{eq:lambdaPlusSol} and~\eqref{eq:nuPlusSol} for $\cos\theta \to 0$.

\bibliographystyle{apsrev}
\bibliography{literature}

\end{document}